\documentclass[twocolumn]{aastex63}

\newcommand{\mjup}{M$_\mathrm{J}$}
\newcommand{\rjup}{R$_\mathrm{J}$}
\newcommand{\msun}{M$_\Sun$}
\newcommand{\rsun}{R$_\Sun$}
\usepackage{textcomp, gensymb, amsmath}

\shorttitle{Tidal Inspiral and Heating in Kepler-1658b}
\shortauthors{Vissapragada et al.}

\begin{document}

\title{The Possible Tidal Demise of Kepler's First Planetary System}

\correspondingauthor{Shreyas~Vissapragada}
\email{shreyas.vissapragada@cfa.harvard.edu}

\author[0000-0003-2527-1475]{Shreyas~Vissapragada}
\altaffiliation{51 Pegasi b Fellow}
\affil{Center for Astrophysics $\vert$ Harvard \& Smithsonian, 60 Garden Street, Cambridge, MA 02138, USA}

\author[0000-0003-1125-2564]{Ashley~Chontos}
\altaffiliation{Henry Norris Russell Fellow}
\affil{Department of Astrophysical Sciences, Princeton University, 4 Ivy Lane, Princeton, NJ 08540, USA}

\author[0000-0002-0371-1647]{Michael~Greklek-McKeon}
\affil{Division of Geological and Planetary Sciences, California Institute of Technology, 1200 East California Blvd, Pasadena, CA 91125, USA}

\author[0000-0002-5375-4725]{Heather~A.~Knutson}
\affil{Division of Geological and Planetary Sciences, California Institute of Technology, 1200 East California Blvd, Pasadena, CA 91125, USA}

\author[0000-0002-8958-0683]{Fei~Dai}
\altaffiliation{NASA Hubble Fellow}
\affil{Division of Geological and Planetary Sciences, California Institute of Technology, 1200 East California Blvd, Pasadena, CA 91125, USA}

\author[0000-0001-7144-589X]{Jorge P\'{e}rez Gonz\'{a}lez}
\affil{Department of Physics \& Astronomy, University College London, Gower Street, WC1E 6BT London, UK}

\author[0000-0003-4976-9980]{Sam~Grunblatt}
\affil{Department of Physics and Astronomy, Johns Hopkins University, 3400 N. Charles Street, Baltimore, MD 21218, USA}

\author[0000-0001-8832-4488]{Daniel~Huber}
\affil{Institute for Astronomy, University of Hawai‘i at M\={a}noa, 2680 Woodlawn Drive, Honolulu, HI 96822, USA}

\author[0000-0003-2657-3889]{Nicholas~Saunders}
\altaffiliation{NSF Graduate Research Fellow}
\affil{Institute for Astronomy, University of Hawai‘i at M\={a}noa, 2680 Woodlawn Drive, Honolulu, HI 96822, USA}

\begin{abstract}
We present evidence of tidally-driven inspiral in the Kepler-1658 (KOI-4) system, which consists of a giant planet (1.1\rjup, 5.9\mjup) orbiting an evolved host star (2.9\rsun, 1.5\msun). Using transit timing measurements from Kepler, Palomar/WIRC, and TESS, we show that the orbital period of Kepler-1658b appears to be decreasing at a rate $\dot{P} = 131_{-22}^{+20}$~ms~yr$^{-1}$, corresponding to an infall timescale $P/\dot{P}\approx2.5$~Myr. We consider other explanations for the data including line-of-sight acceleration and orbital precession, but find them to be implausible. The observed period derivative implies a tidal quality factor $Q_\star' = 2.50_{-0.62}^{+0.85}\times10^4$, in good agreement with theoretical predictions for inertial wave dissipation in subgiant stars. Additionally, while it probably cannot explain the entire inspiral rate, a small amount of planetary dissipation could naturally explain the deep optical eclipse observed for the planet via enhanced thermal emission. As the first evolved system with detected inspiral, Kepler-1658 is a new benchmark for understanding tidal physics at the end of the planetary life cycle.
\end{abstract}

\section{Introduction} \label{sec:intro}
Close-in planets experience intense tidal interactions that can lead to changes in the planetary rotation rate, energy budget, and orbit \citep{Hut1980, Hut1981, Jackson2008, Levrard2009, Leconte2010, Ogilvie2014}. Indeed, the long-term fates of hot Jupiters are thought to be dictated by tides. As tides rob energy from a hot Jupiter's orbit, it spirals in towards its host star, eventually colliding after a few billion years of evolution \citep{Rasio1996, Patzold2004, Levrard2009, Matsumura2010, Schlaufman2013, Hamer2019}. However, these effects are difficult to observe on human timescales, which limits our ability to constrain fundamental tidal parameters that are often uncertain by many orders of magnitude. The only unambiguous example of a tidally decaying orbit so far is that of WASP-12b, where $P/\dot{P}\approx3$~Myr \citep{Maciejewski2016, Patra2017, Maciejewski2018, Baluev2019, Yee2020, Turner2021, Wong2022}. There are a number of other planets that appear to have decaying orbits \citep[recently catalogued by][]{Ivshina2022}; these are worthy of careful scrutiny, but ruling out other astrophysical effects that operate on similar timescales can be challenging \citep{Yee2020, Bouma2020, Maciejewski2021}.

The prospects for observing tidal inspiral may be more favorable for planets orbiting evolved stars. Tides depend sensitively on the inverse scaled semi-major axis $R_\star/a$, so close-in planets around larger stars are natural targets for observing tides in action. Additionally, evolved stars are probably more dissipative than their main sequence counterparts \citep{Villaver2009, Schlaufman2013, Weinberg2017, Barker2020}, so inspiral should be more rapid for their planets to the extent that orbital energy is dissipated in the star. We were therefore motivated to monitor transiting planets on close-in orbits ($P < 5$~days) around subgiant stars.

Kepler-1658b (KOI-4.01) is one such system. This was the first planet candidate revealed by the Kepler mission, as KOI-1.01, KOI-2.01, and KOI-3.01 were known prior to launch \citep{Borucki2011}. Though the planet was mis-classified as a false positive for nearly a decade, \citet{Chontos2019} recently characterized the host star with asteroseismology ($M_\star =$ 1.45\msun; $R_\star =$ 2.89~\rsun) and confirmed the planet with radial velocity (RV) observations ($M_\mathrm{p} =$ 5.88~\mjup; $R_\mathrm{p} =$ 1.07~\rjup). These authors searched the Kepler data for hints of a decaying orbital period, but did not find any on the four~year timescale of the mission. In this work, we present follow-up observations of Kepler-1658b with Palomar/WIRC \citep{Wilson2003} and TESS \citep{Ricker2015}. By extending the observational baseline for this system to 13 years, we were able to search for long-term changes in the orbital architecture that were not previously observable.

\section{Observations} \label{sec:obs}

\subsection{Kepler}
The Kepler spacecraft observed Kepler-1658 for 12 quarters at 30~min cadence and three quarters at 1~min cadence. We downloaded the Kepler light curve using \textsf{lightkurve} \citep{lightkurve} and modeled this dataset using \textsf{exoplanet} \citep{exoplanet:exoplanet}. We fit the light curve quarter-by-quarter to obtain mid-quarter transit timings (defined to be the first transit after the midpoint of the quarter). Except for the mid-quarter time and the limb darkening coefficients, we used the posteriors from the fit in \citet{Chontos2019} as priors for our analysis to inform the model for each quarter of the average transit shape. We simultaneously fit the rotational variability of the star using a Gaussian Process (GP), defining a \textsf{SHOTerm} in \textsf{celerite2} \citep{celerite1, celerite2} with a free amplitude scale $\sigma$, a fixed timescale of 5.66~days (the known rotation period of the star), and a fixed quality factor of 1.

\begin{figure*}
    \centering
    \includegraphics[width=\textwidth]{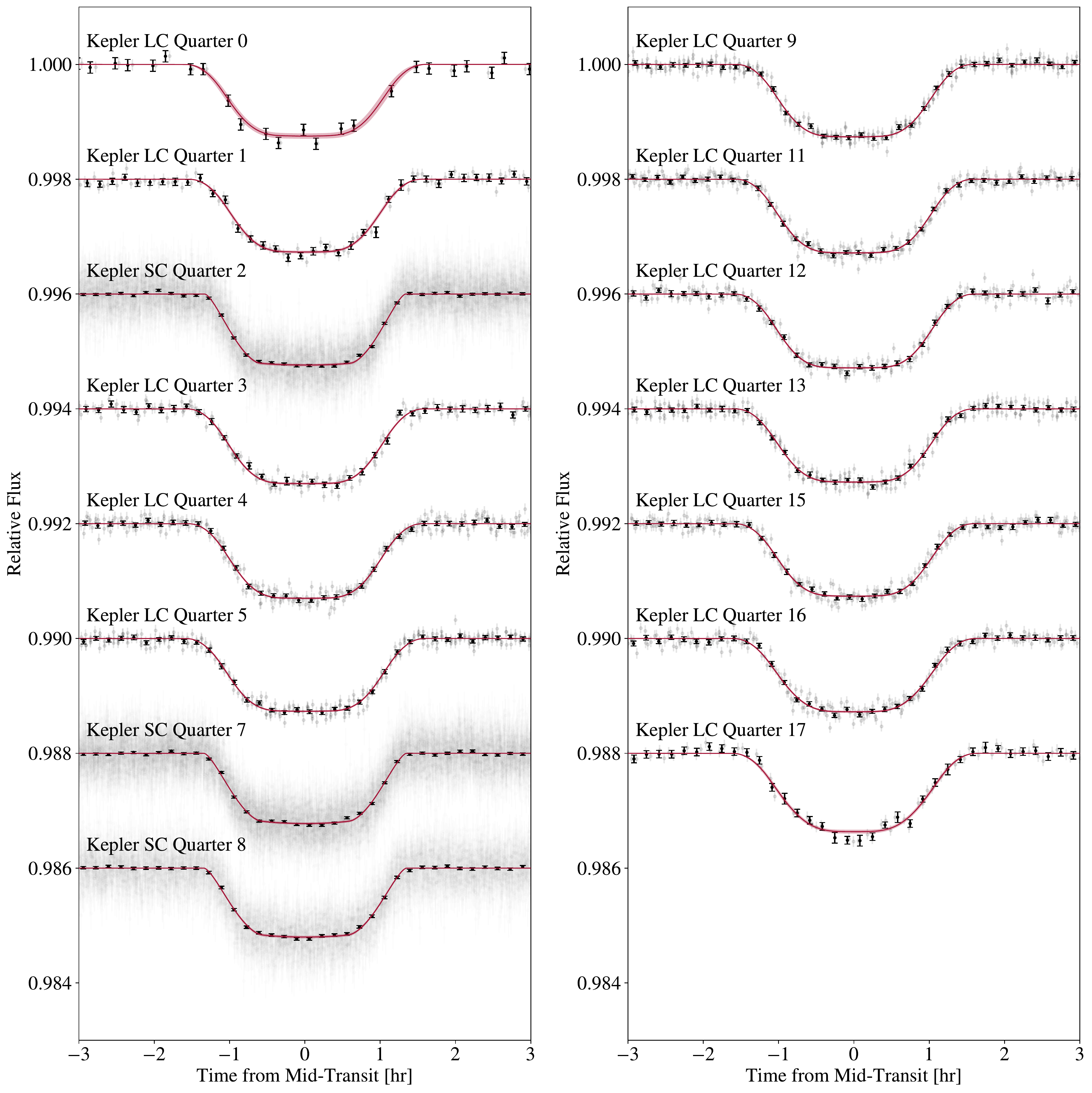}
    \caption{Transit light curves from Kepler. Data are shown binned to 10~minute cadence with the best-fit models given in red.}
    \label{fig:lightcurves_kepler}
\end{figure*}

We ran four chains for all of the fits in this paper. Each chain was tuned for 2000 steps before 1000 posterior draws were taken. For each fit, we verified that the Gelman-Rubin \citep{Gelman1992} statistic was $\hat{R}\ll1.01$ for all sampled parameters. The transit times are given in Table~\ref{table:times}, and the light curves are plotted in Figure~\ref{fig:lightcurves_kepler}.

\begin{deluxetable}{ccccc}[ht]
\tablecaption{Transit times for Kepler-1658b. \label{table:times}}
\tablehead{\colhead{Dataset} & \colhead{Transit Time (BJD$_\mathrm{TDB}$)}}
\startdata
Kepler LC Quarter 0 & $2454959.7314_{-0.0015}^{+0.0014}$ \\
Kepler LC Quarter 1 & $2454982.82835_{-0.00061}^{+0.00061}$ \\
Kepler SC Quarter 2 & $2455048.26751_{-0.00022}^{+0.00021}$ \\
Kepler LC Quarter 3 & $2455140.65189_{-0.00042}^{+0.00040}$ \\
Kepler LC Quarter 4 & $2455233.03736_{-0.00035}^{+0.00033}$ \\
Kepler LC Quarter 5 & $2455325.42133_{-0.00036}^{+0.00035}$ \\
Kepler SC Quarter 7 & $2455510.19192_{-0.00023}^{+0.00023}$ \\
Kepler SC Quarter 8 & $2455602.57708_{-0.00027}^{+0.00027}$ \\
Kepler LC Quarter 9 & $2455691.11211_{-0.00031}^{+0.00033}$ \\
Kepler LC Quarter 11 & $2455883.58121_{-0.00033}^{+0.00032}$ \\
Kepler LC Quarter 12 & $2455975.96583_{-0.00036}^{+0.00034}$ \\
Kepler LC Quarter 13 & $2456064.50087_{-0.00036}^{+0.00036}$ \\
Kepler LC Quarter 15 & $2456256.97026_{-0.00035}^{+0.00037}$ \\
Kepler LC Quarter 16 & $2456349.35438_{-0.00039}^{+0.00038}$ \\
Kepler LC Quarter 17 & $2456410.94385_{-0.00064}^{+0.00065}$ \\
Palomar/WIRC Visit 1 & $2459097.8002_{-0.0015}^{+0.0015}$ \\
TESS Sector 41 & $2459436.5407_{-0.0023}^{+0.0023}$ \\
Palomar/WIRC Visit 2 & $2459790.6819_{-0.0013}^{+0.0015}$ \\
TESS Sector 54 & $2459786.8359_{-0.0030}^{+0.0028}$ \\
TESS Sector 55 & $2459813.7791_{-0.0027}^{+0.0029}$ \\
\enddata
\tablecomments{For the Kepler datasets, LC and SC refer to Long Cadence (30~min exposures) and Short Cadence (1~min exposures), respectively.}
\end{deluxetable}

\begin{figure}
    \centering
    \includegraphics[width=0.42\textwidth]{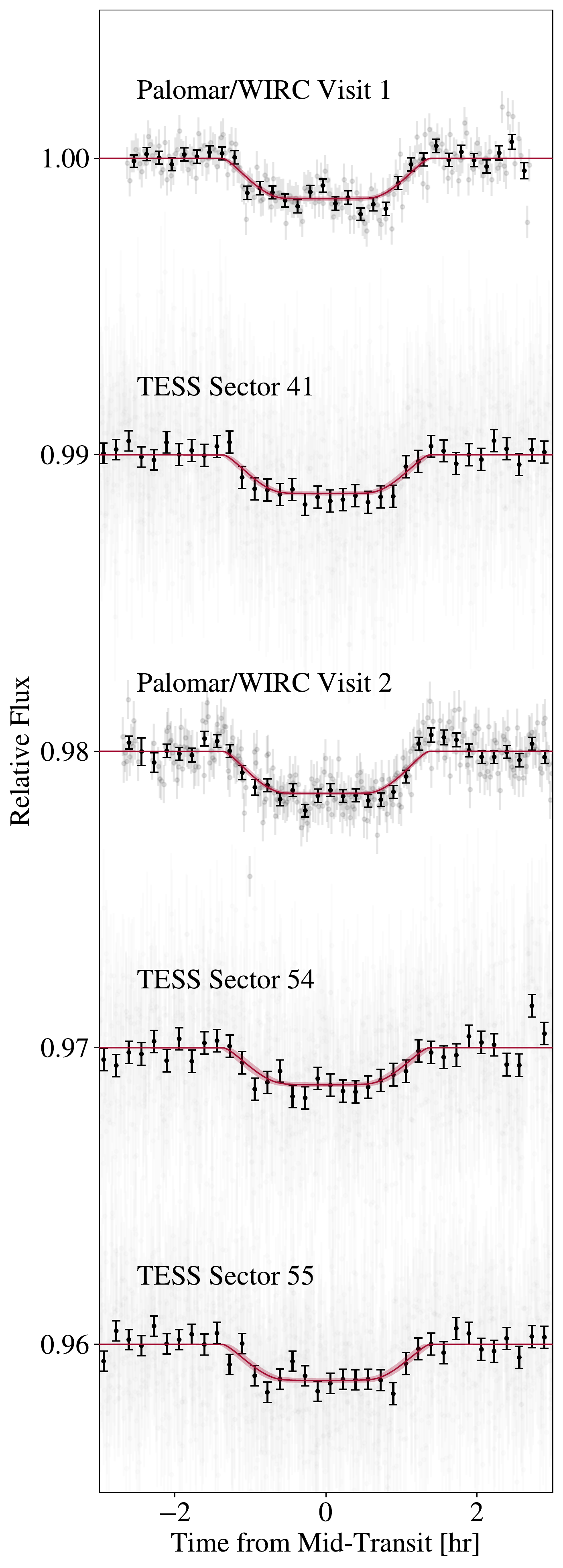}
    \caption{Transit light curves from Palomar/WIRC and TESS. Data are shown binned to 10~minute cadence with the best-fit models given in red.}
    \label{fig:lightcurves_wirc}
\end{figure}

\subsection{Palomar/WIRC}
We observed transits of Kepler-1658b with Palomar/WIRC on UT September 5 2020 and UT July 26 2022. We collected data in $J$ band using a beam-shaping diffuser \citep{Stefansson2017, Vissapragada2020, Greklek-McKeon2022}. On the first night, we acquired 60~second exposures from 04:28 to 09:51 UT (airmass 1.0--2.2), and on the second night, we acquired 40~second exposures from 05:04 to 11:37 UT (airmass 1.1--1.7). Images from both nights were dark-corrected, flat-fielded, and background-subtracted per the methods in \citet{Vissapragada2020} and \citet{Greklek-McKeon2022}. We then performed aperture photometry on Kepler-1658 along with ten comparison stars on the first night and nine on the second night. We used uncontaminated annuli with 25~pixel inner radii and 50~pixel outer radii for local background subtraction. We tested aperture sizes from 5--25~pixels on both nights and chose the apertures that minimized the scatter in the final light curves, which was 18~pixels (4$\farcs$5) for the first night and 12~pixels (3$\farcs$0) for the second night. 

We fit the Palomar/WIRC light curves using \textsf{exoplanet}, again using the results from \citet{Chontos2019} as priors for the fit for all parameters except the mid-transit times and limb darkening coefficients. We modeled the systematics as a linear combination of comparison star light curves and the mean-subtracted times \citep{Vissapragada2020, Greklek-McKeon2022}. We also tried including combinations of additional decorrelation vectors into the systematics model, including the PSF centroid offset, the background level of the image, and the airmass. We fit the light curve with all subsets of these three vectors and chose the systematics model that minimized the Bayesian Information Criterion (BIC). For the first night, the BIC was minimized when including only the background level, whereas on the second night the BIC was minimized when using only the airmass. The light curves are shown in Figure~\ref{fig:lightcurves_wirc}, and the transit times are reported in Table~\ref{table:times}.

\subsection{TESS}
TESS obtained photometry for Kepler-1658 (TOI-4480.01) in sectors 41, 54, and 55, all at 2~min cadence. We fit the TESS photometry sector-by-sector using the \citet{Chontos2019} priors for all values except the mid-sector transit time and the limb darkening coefficients. To handle the stellar variability in the TESS bandpass, we used the same GP parameters from the Kepler fit. The results are shown in Figure~\ref{fig:lightcurves_wirc}, and the transit times are reported in Table~\ref{table:times}. To ensure our results were robust to choices in modeling methodology, three of us (F. D., M. G.-M., N. S.) reduced and fit the TESS data using independent pipelines and obtained consistent results.
\begin{figure*}
    \centering
    \includegraphics[width=\textwidth]{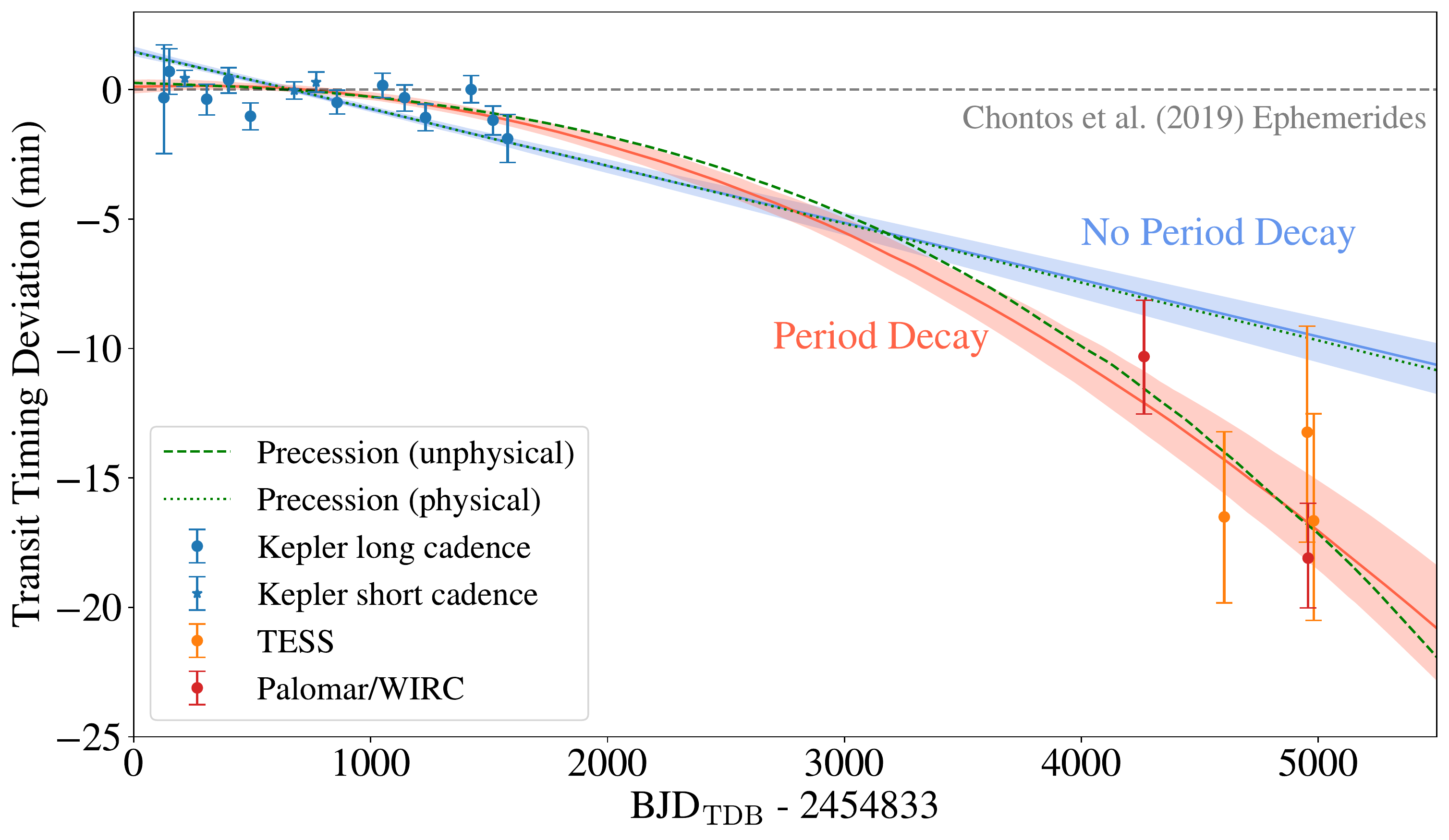}
    \caption{Transit timing data for Kepler-1658b relative to the ephemerides from \citet{Chontos2019}. The Kepler data (blue points) are consistent with the original ephemerides, but the Palomar/WIRC (red points) and TESS (orange points) data are not. The model including a decaying orbital period (orange curve) fits the new data better than the best-fit linear ephemerides (blue curve). While models allowing for an arbitrary precession rate (dashed green curve) match the data as well as the period decay model, the maximum physically-allowed precession rate from Equation (\ref{maxprecession}; dotted green curve) cannot improve the fit over the nominal model.}
    \label{fig:decay}
\end{figure*}

\section{Results} \label{sec:res}

\subsection{Fitting the Transit Times}
The transit times are shown in Figure~\ref{fig:decay} with the ephemerides from \citet{Chontos2019} subtracted off. In agreement with their work, we found no compelling evidence for a changing orbital period in the Kepler data alone. However, the transits observed by Palomar/WIRC and TESS arrived early. In light of these new data, we sought to quantify the evidence for an evolving orbital period.

We fit the transit times as a function of observing epoch $t(N)$ using two models \citep{Yee2020}. The first model fit the data with a constant orbital period:
\begin{equation}
    t(N) = t_0 + NP,
    \end{equation}
where $t_0$ was the transit time at the reference epoch from \citet{Chontos2019}. The second model fit the data including a constant period derivative $dP/dN$:
\begin{equation}
    t(N) = t_0 + NP + \frac{1}{2}\frac{dP}{dN}N^2.
\end{equation}
We used the nested sampling tool \textsf{dynesty} \citep{Speagle2020} to estimate the posteriors and Bayesian evidences $\mathcal{Z}$ for these models. We performed each nested sampling run using single ellipsoid bounds, 1000 live points, and the random walk sampling method, terminating each run when the estimated log-evidence remaining was less than $0.01$. The priors, posteriors, and evidences are given in Table~\ref{table:modelselection}. We then computed the Bayes factor $\ln B = \ln\mathcal{Z}_2 - \ln\mathcal{Z}_1 = 17.5$ for this model comparison, representing decisive evidence \citep{Trotta2008} for the period derivative model over the constant period model. We conclude that the orbital period of the planet appears to be decreasing at a rate of 131$^{+20}_{-22}$ ms~yr$^{-1}$.

\begin{deluxetable*}{ccccc}[t]
\tablecaption{Model selection for the timing data. \label{table:modelselection}}
\tablehead{\colhead{Model} & \colhead{Parameter} & \colhead{Unit} & \colhead{Prior} & \colhead{Posterior}}
\startdata
No Decay & $t_0$ & BJD$_\mathrm{TDB}$ & $\mathcal{U}(t_\mathrm{c} - 1, t_\mathrm{c} + 1)$ & $2455005.92478_{-0.00014}^{+0.00013}$\\
 $\log(\mathcal{Z})$ = 84.3 & $P$ & days & $\mathcal{U}(P_\mathrm{c} - 0.001, P_\mathrm{c} + 0.001)$ & $3.84936720_{-0.00000066}^{+0.00000060}$ \\
 \hline
Decay & $t_0$ & BJD$_\mathrm{TDB}$ & $\mathcal{U}(t_\mathrm{c} - 1, t_\mathrm{c} + 1)$ & $2455005.92415_{-0.00016}^{+0.00017}$ \\
  $\log(\mathcal{Z})$ = 101.8 & $P$ & days & $\mathcal{U}(P_\mathrm{c} - 0.001, P_\mathrm{c} + 0.001)$ & $3.8493733_{-0.0000011}^{+0.0000012}$ \\
 & $\log_{10}(-dP/dN)$ & $\log_{10}$(days~orbit$^{-1}$) & $\mathcal{U}(-10, -6)$ & $-7.796_{-0.079}^{+0.061}$ \\
 \hline
 Precession & $t_0$ & BJD$_\mathrm{TDB}$ & $\mathcal{U}(t_\mathrm{c} - 1, t_\mathrm{c} + 1)$ & $2455005.8771_{-0.0061}^{+0.0086}$\\
  $\log(\mathcal{Z}) = 98.9$ & $P_\mathrm{s}$ & days & $\mathcal{U}(P_\mathrm{c} - 0.001, P_\mathrm{c} + 0.001)$ & $3.8493936_{-0.0000072}^{+0.0000085}$ \\
 & $e\cos\omega_0$ & & $\mathcal{N}$(-0.00840, 0.00080) & $-0.00836_{-0.00076}^{+0.00076}$\\
 & $e\sin\omega_0$ & & $\mathcal{N}$(0.062, 0.019) & $0.059_{-0.022}^{+0.021}$ \\
 & $\log_{10}(d\omega/dN)$ & $\log_{10}$(rad~orbit$^{-1}$) & $\mathcal{U}(-8, 2)$ & $-3.159_{-0.071}^{+0.085}$ \\
\enddata
\tablecomments{$\mathcal{U}(a,b)$ indicates a uniform prior with lower bound $a$ and upper bound $b$. $\mathcal{N}(a,b)$ indicates a normal prior with mean $a$ and standard deviation $b$. $P_c$ and $t_c$ are shorthand for the best-fit ephemerides from \citet{Chontos2019}.}
\end{deluxetable*}

\subsection{Considering Alternate Explanations}
There are a number of effects that can cause apparent decreases in the planetary orbital period on decade-long timescales \citep[e.g.][]{Patra2017, Yee2020, Bouma2020, Maciejewski2021, Ivshina2022}. Only when these effects are ruled out can we definitively attribute our observations to orbital decay.
\subsubsection{Line-of-Sight Acceleration}
We first consider line-of-sight acceleration effects: perhaps the transit times are arriving earlier than expected because the system is accelerating towards us along our line of sight. If it did cause the observed period derivative for Kepler-1658b, such an acceleration would manifest as a linear trend in the RV data for this planet, with magnitude $\dot{v}_r = c\dot{P}/P = -0.324_{-0.049}^{+0.054}$~m~s$^{-1}$d$^{-1}$. We searched for this acceleration in the RV data from \citet{Chontos2019}. We re-fit the RVs in \textsf{exoplanet} accounting for an acceleration term, but we found it to be consistent with zero: $\dot{v}_r = +0.047_{-0.067}^{+0.069}$~m~s$^{-1}$d$^{-1}$. At a confidence of 4.3$\sigma$, we conclude that line-of-sight acceleration cannot explain the early transit times observed by Palomar/WIRC and TESS. 

\subsubsection{Apsidal Precession}
Next, we consider apsidal precession of an eccentric orbit, which can mimic decay due to tides if the full precession cycle is not captured \citep[e.g.][]{Patra2017, Yee2020}. Kepler-1658b's orbit \textit{is} slightly eccentric, as its secondary eclipse arrives about half an hour early relative to an orbital phase of 0.5 in the Kepler data \citep{Chontos2019}, so we must consider this scenario carefully. We fit the data with a constant precession rate using the model from \citet{Gimenez1995}, used previously by \citet{Patra2017, Yee2020}:
\begin{align}
    t(N) &= t_0 + NP_\mathrm{s} - \frac{eP_\mathrm{a}}{\pi}\cos\omega(N),\\
    \omega(N) &= \omega_0 + \frac{d\omega}{dN}N \nonumber \\
    P_\mathrm{s} &= P_\mathrm{a}\Big(1 - 2\pi\frac{d\omega}{dN}\Big), \nonumber
\end{align}
where $d\omega/dN$ is the precession rate and $P_\mathrm{s}$ and $P_\mathrm{a}$ are the sidereal and anomalistic periods, respectively. The priors, posteriors, and evidence for this model are given in Table~\ref{table:modelselection}. We find that precession is capable of fitting the data just as well as the period decay model (Figure~\ref{fig:decay}), with relatively insignificant evidence $\ln B = 2.9$ in favor of the decay model. The required precession rate $d\omega/dN \approx 7\times10^{-4}$~rad~orbit$^{-1}$ is about $4 \degree$~yr$^{-1}$. 

However, this rate is severely problematic from a theoretical standpoint. If the precession is due to the planetary tidal bulge, the rate of precession constrains the planet's Love number $k_{2\mathrm{p}}$ \citep{Ragozzine2009, Patra2017}:
\begin{align}
    \frac{d\omega}{dN} &= 15\pi k_{2\mathrm{p}}\Big(\frac{M_\star}{M_\mathrm{p}}\Big)\Big(\frac{R_\mathrm{p}}{a}\Big)^5 \\
    &\approx (8.9_{-2.1}^{+2.6}\times10^{-7}~\mathrm{rad~orbit^{-1}})k_{2\mathrm{p}}. \nonumber
\end{align}
The Love number is related to internal structure and ranges from 0 to 1.5 \citep[e.g.][]{Russell1928, Sterne1939}, but matching the required precession rate requires an unphysical $k_{2\mathrm{p}} \sim 10^{3}$ for the planet. The precession rate from the planetary tidal bulge is not physically permitted to exceed values of order $\sim10^{-6}$~rad~orbit$^{-1}$.

For hot Jupiters orbiting evolved stars, the stellar tidal bulge can be more important \citep{Ragozzine2009}:
\begin{align}
    \frac{d\omega}{dN} &= 15\pi k_{2\star}\Big(\frac{M_\mathrm{p}}{M_\star}\Big)\Big(\frac{R_\mathrm{\star}}{a}\Big)^5 \label{maxprecession} \\
    &\approx (1.70_{-0.35}^{+0.45}\times10^{-4}~\mathrm{rad~orbit^{-1}})k_{2\star}. \nonumber
\end{align}
Though precession from the stellar bulge is much faster, an unphysical $k_{2\star} \approx 4$ is still required. Precession induced by the rotational flattening of the star and planet are even smaller contributions than the tidal bulge terms, and the precession rate from general relatively is two orders of magnitude too small as well \citep{Ragozzine2009}.

Finally, we consider the potential impact of an unseen outer body in the system. Secular perturbations from the outer body with mass $M_\mathrm{o}$, semimajor axis $a_\mathrm{o}$ and eccentricity $e_\mathrm{o}$ would drive precession of the inner planet; in the limit of an apsidally anti-aligned outer body with a large eccentricity, the precession rate can be written approximately as \citep{Mardling2007, Batygin2009}:
\begin{align}
    \frac{d\omega}{dN} &= \frac{15}{8}\pi\Big(\frac{M_\mathrm{o}}{M_\star}\Big)\Big(\frac{a}{a_\mathrm{o}}\Big)^4\frac{e_0}{e(1 - e_\mathrm{o}^2)^{5/2}} \\
    &= (5.67_{-0.36}^{+0.38}\times10^{-6}~ \mathrm{rad~orbit^{-1}})\Big(\frac{M_\mathrm{o}}{10\mathrm{M}_\mathrm{J}}\Big)\Big(\frac{a_\mathrm{o}}{1~\mathrm{au}}\Big)^{-4}\nonumber \\
    &\times \frac{e_\mathrm{o}}{(1 - e_\mathrm{o}^2)^{5/2}} \nonumber
\end{align}
To match the necessary precession rate, an outer body with $10$\mjup~at $1$~au would need an eccentricity of $e\approx0.9$. A body with this mass, separation, and eccentricity would have been readily observed in the RVs, with semiamplitude $(540~$m~s$^{-1}$)$\sin i$, but the data do not indicate its presence. Unless the inclination of the outer body is very close to $i = 0\degree$ (requiring some fine-tuning), this mechanism also seems unlikely.

We conclude that, while apsidal precession model can match the data, it requires an implausibly large precession rate and can thus be ruled out on physical grounds. Still, it would be helpful to strengthen this conclusion further with observations. For WASP-12b, the decisive evidence against apsidal precession came from the timing of secondary eclipses, which should arrive late for a precessing orbit \citep{Patra2017, Yee2020}. However, because the eclipse is quite shallow for Kepler-1658b ($62$~ppm in the Kepler bandpass, and ostensibly $200-300$~ppm in the near-infrared), it is difficult to make the same measurement from the ground or with TESS. Now that we are a decade removed from the Kepler era, a secondary eclipse observation with HST or JWST would be helpful for testing the precession model more definitively. 

\section{Discussion}
\label{sec:disc}

\subsection{Tidal Decay}
The best explanation for the early transit times observed by Palomar/WIRC and TESS is tidal decay of Kepler-1658b's orbit. Given our observed $\dot{P}$, we can calculate the rate at which orbital energy is being lost \citep[e.g.][]{Millholland2018}:
\begin{equation}
    \frac{dE}{dt} = \frac{(GM_\star)^{3/2}M_p}{6\pi}a^{-5/2}\dot{P} = -1.08^{+0.20}_{-0.21}\times10^{31}~\mathrm{erg}~\mathrm{s}^{-1} \label{energy}
\end{equation}

To estimate the implied tidal quality factors, we assume the constant time lag model of \citet{Leconte2010} with zero stellar and planetary obliquity \citep[though both could feasibly increase the dissipation if maintained over long timescales, ][]{Leconte2010, Millholland2018, Su2022}. As the eccentricity is small, it is acceptable to truncate the model at second order in $e$ \citep[Equation 22 in][]{Leconte2010}. Then, for inspiral dominated by dissipation in the star, the implied stellar tidal quality factor is:
\begin{align}
     Q_\star' &= \Big(\frac{M_\mathrm{p}}{M_\star}\Big)\Big(\frac{R_\star}{a}\Big)^5 \frac{27\pi}{\dot{P}}\Big(\frac{\omega_\star}{n} - 1\Big) \\
    &=2.50_{-0.62}^{+0.85}\times10^4. \nonumber
\end{align}
Dynamical tides are required to make the star so dissipative. For WASP-12 \citep[with $Q'_\star\sim10^5$;][]{Yee2020, Turner2021, Wong2022}, it has been suggested that $g$ modes deposit their energy efficiently via wave-breaking in the radiative core \citep{Weinberg2017, Bailey2019, Barker2020}. This explanation is inadequate for Kepler-1658, for which the \citet{Barker2020} models predict $Q'_\star \sim 10^8$ with this mechanism. Dissipation of inertial waves in the convective zone \citep{Ogilvie2007} are more effective for this rapidly-rotating star. For their model closest to Kepler-1658, \citet{Barker2020} find that inertial wave dissipation should result in $Q'_\star \sim 10^4$, in good agreement with our observations. If stellar dissipation drives the inspiral of Kepler-1658b, inertial waves are likely responsible.

If instead the inspiral is dominated by dissipation in the planet, then assuming the planet is tidally locked, the implied planetary tidal quality factor is:
\begin{align}
    Q_\mathrm{p}' &= -\frac{3}{2}\Big(\frac{M_\star}{M_\mathrm{p}}\Big)\Big(\frac{R_\mathrm{p}}{a}\Big)^5 \frac{171\pi e^2}{\dot{P}} \\
    &= 13.5^{+4.9}_{-3.5}\Big(\frac{e}{0.06}\Big)^2. \nonumber
\end{align}
A similar expression is given by \citet{Efroimsky2022}, who found planetary eccentricity tides to be capable of explaining the orbital decay of WASP-12b. For Kepler-1658b, we require $Q_\mathrm{p}'\sim10$ if all the energy is dissipated in the planet. The quality factors for Jupiter and Saturn are much larger, $Q_\mathrm{p}'\sim10^{5}$ \citep{Goldreich1966}, but dynamical tides can lead to small effective quality factors via $f$ mode diffusive growth and/or gravity wave dissipation in the radiative upper envelope of the planet \citep{Lubow1997, Ogilvie2004, Wu2018}. The former mechanism cannot work as it requires a pericenter distance of less than four tidal radii (about 5\rsun~for this system), but it seems plausible that the outer part of the planet is radiative and could support the resonant excitation of $g$ modes.

Given these estimates for the tidal quality factors, the majority of the energy is probably lost in the star and not the planet. But even a small amount of planetary dissipation would have observable consequences for the planetary energy budget as we describe in the next section. Finally, it is worth noting that the stellar spin period (5.66~days) and planetary orbital period (3.85~days) are close to a near-integer ratio, with $\omega_\star / n \approx 2/3$. If this is not coincidental, the near-commensurability may encode the system's history of tidal angular momentum exchange.

\subsection{Tidal Superheating}
A secondary eclipse has been detected for Kepler-1658b in the optical, with depth $62\pm4$~ppm \citep{Chontos2019}. If this eclipse is due solely to reflected light, the geometric albedo would be $A_\mathrm{g} = 0.72 \pm 0.09$ -- by far the largest ever measured for a hot Jupiter. This is comparable to the geometric albedos of icy satellites in the Solar System \citep{Madden2018} and is unexpected for hot Jupiters \citep[e.g.][]{Adams2022}. We find it more likely that the planet is over-luminous in the optical due to its own thermal emission, but the maximum dayside temperature of this planet is $T_\mathrm{maximum}=2796\pm73$~K \citep{Cowan2011}. At this blackbody temperature, the thermal contribution in the Kepler bandpass is negligible at $\lesssim 15$~ppm; the dayside must be $T_\mathrm{observed}\approx3450$~K to match the eclipse depth via thermal emission. The dayside temperature of the planet cannot be due to stellar irradiation alone. 

The extra energy required to superheat the planet beyond its maximum dayside temperature could be provided by dissipation in the planet. This requires the luminosity from eccentricity tides to be a substantial fraction of the incident stellar irradiation \citep[e.g.][]{Bodenheimer2001, Jackson2008}. To superheat the dayside of the planet requires an additional luminosity:
\begin{equation}
    \Delta L = \sigma_\mathrm{SB}\pi R_\mathrm{p}^2(T_\mathrm{observed}^4 - T_\mathrm{maximum}^4) \approx 8\times10^{29}~\mathrm{erg~s}^{-1}
\end{equation} This is an order of magnitude smaller that the total rate at which orbital energy is being dissipated (Equation~\ref{energy}); it is plausible that $\lesssim10\%$ of the energy from the shrinking orbit is being dissipated in the planet. Planetary dissipation is especially interesting considering that many gas giants orbiting evolved stars appear to be (re-)inflated, which requires the deposition of additional energy beyond the incident stellar radiation \citep{Grunblatt2016, Lopez2016, Grunblatt2017}. If a similar mechanism operates for these systems, tidal heating could be a natural explanation for the inflation of planets orbiting evolved stars.

\section{Conclusion} \label{sec:conc}
Using data from Kepler, Palomar/WIRC, and TESS, we showed that Kepler-1658b's orbit appears to be shrinking at a rate of $\dot{P} = 131_{-22}^{+20}$~ms~yr$^{-1}$, corresponding to an inspiral timescale of $P/\dot{P}\approx$~2.5~Myr. We ruled out line-of-sight-acceleration effects using RVs, and found that apsidal precession could not explain the data either, as the required precession rates were unphysical. Dissipation in the star is the most likely culprit: our inspiral rate corresponds to a modified stellar tidal quality factor $Q_\star = 2.50_{-0.62}^{+0.85}\times10^4$, which agrees well with models of dynamical tides invoking inertial wave dissipation \citep{Barker2020}. Planetary dissipation probably cannot explain the entire inspiral rate, but we found it plausible that some ($\lesssim10\%$) of the energy from the shrinking orbit is being dissipated in Kepler-1658b itself, which would explain its apparently superheated dayside.

We encourage continued transit observations of this system, as they will help improve the constraint on $\dot{P}$. Additionally, a secondary eclipse observation of this system at thermal wavelengths would simultaneously clarify the dayside temperature of the planet and definitively test the orbital precession hypothesis. It would also be helpful to constrain the stellar obliquity via the Rossiter-McLaughlin effect or Doppler Tomography for this rapidly-rotating star ($v\sin i_\star = 34$~km~s$^{-1}$). Our tidal calculations also neglected the role of both planetary and stellar obliquity, but these could help drive the orbital decay \citep{Leconte2010, Millholland2018}. 

Finally, many new planets orbiting evolved stars are being discovered with TESS \citep{Grunblatt2022, Grunblatt2022b, Saunders2022}. If the tidal quality factor obtained here is applicable to other evolved planet-hosting stars, then most of their planets are nearing the ends of their lives \citep{Schlaufman2013, Hamer2019}, and we should begin to see hints of orbital decay for these planets within the next decade. The growing population of planets orbiting evolved stars is an exciting new laboratory for many of the ideas we have presented here.

\acknowledgments
We thank the Palomar Observatory telescope operators and support astronomers for their support of this work. We additionally thank Adrian Barker, Konstantin Batygin, Dave Charbonneau, Jim Fuller, Mercedes Lopez-Morales, Morgan MacLeod, and Sam Yee for insightful comments and discussions.

This paper is based on data collected by the TESS mission. Funding for the TESS mission is provided by NASA's Science Mission Directorate. We acknowledge the use of public TESS data from pipelines at the TESS Science Office and at the TESS Science Processing Operations Center. This research has made use of the Exoplanet Follow-up Observation Program website, which is operated by the California Institute of Technology, under contract with the National Aeronautics and Space Administration under the Exoplanet Exploration Program. D. H. acknowledges support from the Alfred P. Sloan Foundation and the National Aeronautics and Space Administration (80NSSC19K0597, 80NSSC21K0652).

\facilities{ADS, NASA Exoplanet Archive, Kepler, Hale 200-inch, TESS}
\software{\textsf{exoplanet} \citep{exoplanet:exoplanet}, \textsf{lightkurve} \citep{lightkurve}, \textsf{pymc3} \citep{exoplanet:pymc3}, \textsf{celerite2} \citep{celerite1, celerite2}, \textsf{arviz} \citep{exoplanet:arviz}, \textsf{astropy} \citep{exoplanet:astropy13, exoplanet:astropy18}, \textsf{dynesty} \citep{Speagle2020} }
\clearpage
\bibliography{references}
\end{document}